\documentclass[a4paper,amsfonts,amssymb,amsmath,reprint,nofootinbib,twoside,superscriptaddress]{revtex4-1}
\usepackage[english]{babel}
\usepackage[utf8]{inputenc}
\usepackage{graphicx}
\usepackage{microtype}
\usepackage{tikz}
\usetikzlibrary{arrows.meta,calc}
\usepackage[hidelinks]{hyperref}
\hypersetup{pdftitle={Analytic response functions of space-based gravitational-wave detectors to ultralight scalar-field dark matter under multiple types of TDI combinations},
  pdfauthor={Wenqiu Tang, Pan-Pan Wang, Yuezhe Chen, Rui Luo, Cheng-Gang Shao}}
\graphicspath{{figures/}{images/}}
\numberwithin{equation}{section}

\begin{document}
\title{Analytic response functions of space-based gravitational-wave detectors to ultralight scalar-field dark matter under multiple types of TDI combinations}
\author{Wenqiu Tang}
\affiliation{National Gravitation Laboratory, MOE Key Laboratory of Fundamental Physical Quantities Measurement,
School of Physics, Huazhong University of Science and Technology, Wuhan 430074, People’s Republic of China}
\author{Pan-Pan Wang}
\affiliation{College of Physics, Chongqing University, Chongqing 401331, China}
\author{Yuezhe Chen}
\affiliation{National Gravitation Laboratory, MOE Key Laboratory of Fundamental Physical Quantities Measurement,
School of Physics, Huazhong University of Science and Technology, Wuhan 430074, People’s Republic of China}
\author{Rui Luo}
\affiliation{National Gravitation Laboratory, MOE Key Laboratory of Fundamental Physical Quantities Measurement,
School of Physics, Huazhong University of Science and Technology, Wuhan 430074, People’s Republic of China}
\author{Cheng-Gang Shao}
\affiliation{School of Physics and Optoelectronic, Yangtze University, Jingzhou 434023, China}

\begin{abstract}
Ultralight scalar fields are important candidates for dark matter. As coherently oscillating fields, they can leave characteristic signatures in space-based gravitational-wave detectors through their coupling to test masses. In this paper, we derive analytic expressions for the response functions of space-based gravitational-wave detectors to ultralight scalar-field dark matter under multiple types of time-delay interferometry (TDI) combinations. Starting from the coupling action between the scalar field and a test mass, we obtain the additional displacement of the test mass and the single-link Doppler response. We then give the general analytic form of the response function for an arbitrary TDI combination and compare the analytic and Monte Carlo numerical results for the optimal combinations A, E, and T. Finally, we analyze the responses of multiple types of TDI combinations, classify them into ordinary and $\zeta$ types according to their low-frequency behavior, and give the low-frequency response expressions of the individual combinations. The results can be used to evaluate the sensitivity of space-based gravitational-wave missions such as LISA, Taiji, and TianQin to the mass and coupling parameters of ultralight scalar-field dark matter.
\end{abstract}

\maketitle

\section{Introduction}
\label{sec:1}

Dark matter (DM) is one of the major open problems in contemporary physics and astronomy. Galactic rotation curves provide evidence for an invisible matter component~\cite{Rubin1970}, while cosmological observations support a substantial non-baryonic matter density~\cite{Aghanim2020}. Within the base $\Lambda$CDM model, DM accounts for approximately 26\% of the total cosmic energy density~\cite{Aghanim2020}. Its microscopic nature remains unknown.

Complementary searches for particle DM include nuclear-recoil measurements~\cite{Aprile2023,Aalbers2023,Meng2021} and searches for annihilation or decay products~\cite{Ackermann2015}. Ultralight scalar fields (ULSFs) motivate a different approach: their extremely large phase-space occupation allows a classical oscillating-field description~\cite{Marsh2016,Hui2017,Hui2021,Ferreira2021}. A linear coupling to the test-mass rest mass produces an oscillatory force, making precision interferometry a probe of this interaction~\cite{Piazza2010,Graham2016,Morisaki2019}.

Space-based gravitational-wave detectors measure relative test-mass motion through long-baseline laser interferometry. Missions such as LISA, Taiji, and TianQin target low-frequency gravitational-wave observations~\cite{AmaroSeoane2017,Luo2016,Hu2017}. Their displacement measurements are also sensitive to non-gravitational forces associated with scalar DM. The resulting motion produces Doppler shifts in the laser links~\cite{Yu2023}, complementing interferometric and precision-measurement searches~\cite{Pierce2018,Grote2019,Michimura2020,Morisaki2021,Centers2021,Miller2023,Fukusumi2023,Kim2023}.

In an ideal equal-arm Michelson interferometer, common laser frequency fluctuations cancel in the differential readout. Unequal arms in a space-based detector sample those fluctuations at different retarded times, leaving residual laser noise in a simple arm difference. Time-delay interferometry (TDI) combines delayed one-way measurements to suppress this noise~\cite{Tinto1999,Armstrong1999,Estabrook2000,Tinto2021}. Cancellation depends on the delay model and its accuracy. Residual secondary noises must also be propagated through each combination.

Previous work has studied scalar-field responses, velocity effects, and sensitivity estimates for elementary TDI combinations~\cite{Yu2023}. Geometric TDI provides a catalogue of 12-link, 14-link, and 16-link combinations and their delay-polynomial decompositions~\cite{Wang2023}. Building on these results, we derive analytic angular averages of the scalar response and verify them using independent numerical calculations. We then compare the 45 geometric combinations under a common static equal-arm model and noise prescription. The comparison distinguishes response equivalence from sensitivity equivalence and identifies where the low-frequency asymptotes apply.

The paper is organized as follows. Section~\ref{sec:2} establishes the signal model of ultralight scalar-field dark matter in space-based gravitational-wave detectors and derives the single-link response. Section~\ref{sec:3} gives the analytic TDI response expressions and their numerical verification. Section~\ref{sec:4} compares the responses and sensitivities of different TDI combinations and discusses their application to coupling estimates. Section~\ref{sec:5} summarizes the paper. We adopt natural units with $c=\hbar=1$ and restore $c$ where lengths and frequencies are given in SI units.

\section{Principles of dark matter detection with space-based gravitational-wave detectors}
\label{sec:2}

To connect the scalar-field interaction to a measurable signal, we first specify the coupling to matter using the effective Lagrangian. We then derive the motion induced in a test mass by the oscillating field and use the resulting displacement to calculate the Doppler response of a single laser link.

\subsection{Action of dark matter}

We consider an ultralight scalar field as a dark matter candidate and adopt a description in which $\phi$ couples linearly to Standard Model fields. The effective Lagrangian in the low-energy limit is \cite{Kawasaki2013,Marsh2016,Hui2017,Hui2021,Ferreira2021}:
\begin{equation}
\label{L}
\begin{aligned}
\mathcal{L}=\mathcal{L}_\phi+\mathcal{L}_{SM}+\mathcal{L}_{\phi-SM}\,.
\end{aligned}
\end{equation}
Here $\mathcal{L}_{SM}$ is the Standard Model Lagrangian, $\mathcal{L}_\phi$ describes a scalar field of mass $m_\phi$, and $\mathcal{L}_{\phi-SM}$ describes its interactions with matter. The latter two terms are given by
\begin{equation}\mathcal{L}_\phi=-\frac{1}{2}\partial_\mu\phi\partial^\mu\phi-\frac{1}{2}m_\phi^2\phi^2\,,\end{equation}
and
\begin{equation}
\begin{split}
\mathcal{L}_{\phi-SM}&=\kappa \phi \Biggl[ \frac{d_e}{4e^2} F_{\mu\nu} F^{\mu\nu} - \frac{d_g \beta_3}{2g_3} G^A_{\mu\nu} G^{A\mu\nu} \\
&\quad - \sum_{i=e,u,d} (d_{m_i} + \gamma_{m_i} d_g) m_i \bar{\psi}_i \psi_i \Biggr]\,,
\end{split}
\end{equation}
where $\kappa\equiv\sqrt{4\pi G}=\sqrt{4\pi}/M_P$, $G$ is Newton\textquotesingle s constant, and $M_P=G^{-1/2}$ is the Planck mass. The tensors $F_{\mu\nu}$ and $G^A_{\mu\nu}$ describe the electromagnetic and gluon fields, respectively, with $A$ a gluon color index. Here $e$ and $g_3$ are the electromagnetic and strong gauge couplings, $\beta_3$ is the beta function of quantum chromodynamics (QCD), and $\gamma_{m_i}$ are the fermion mass anomalous dimensions. The dimensionless coefficients $d_e$, $d_g$, and $d_{m_i}$ parameterize the scalar couplings to these fields and to the fermion mass terms. The fields $\psi_i$, with masses $m_i$, represent the electron and the $u$ and $d$ quarks, and $\bar\psi_i$ are their Dirac adjoints. The quark couplings can also be expressed as
\begin{equation}
\begin{aligned}
d_{\hat{m}}&=\frac{d_{m_d}m_d+d_{m_u}m_u}{m_d+m_u},\\
d_{\delta m}&=\frac{d_{m_d}m_d-d_{m_u}m_u}{m_d-m_u}\,.
\end{aligned}
\end{equation}
$d_{\hat{m}}$ denotes the coupling to the average quark mass, and $d_{\delta m}$ denotes the coupling related to the quark mass difference. For a macroscopic test mass, the scalar coupling enters through the field dependence of its mass $m(\phi)$. This field dependence of $m(\phi)$ determines the motion of the test mass in the scalar field.

\subsection{Principle of dark matter detection with space-based gravitational-wave detectors}

The classical scalar field admits a plane-wave decomposition in the free-field approximation. To calculate the response for a fixed field velocity, we consider one monochromatic mode,
\begin{equation}\phi\left(t,\vec{x}\right)=\phi_{\vec{k}}e^{i\left(\omega t-\vec{k}\cdot\vec{x}+\theta_0\right)}\,,\end{equation}
where $\phi_{\vec{k}}$ is the amplitude, $\omega$ the angular frequency, $\vec{k}$ the momentum vector, and $\theta_0$ the initial phase (whose value does not affect the discussion and can be set to zero). When the scalar field constitutes Galactic dark matter, its velocity dispersion is $v\sim{10}^{-3}$, corresponding to non-relativistic fluctuations with $\vec{k}\approx m_\phi\vec{v}$ and $\omega\approx m_\phi$. The test-mass response to this field follows from the action
\begin{equation}S=-\int{m\left(\phi\right)\sqrt{-\eta_{\mu\nu}dx^\mu d x^\nu}}\,.\end{equation}

We neglect gravitational effects and describe the test mass by a field-dependent mass $m(\phi)$. The Minkowski metric $\eta_{\mu\nu}$ has signature $(-,+,+,+)$. Greek indices label spacetime components and $i$ labels a spatial component. To derive the equation of motion, we use $dx^\mu=(dt,d\vec{x})$ and temporarily denote the test-mass velocity by $\vec{v}=d\vec{x}/dt$, with $v^2=\vec v\cdot\vec v$. The proper-time element then satisfies
\begin{equation}-\eta_{\mu\nu}dx^\mu dx^\nu=dt^2(1-v^2)\,.\end{equation}
Writing the action as $S=\int L\,dt$ gives the point-particle Lagrangian
\begin{equation}L=-m(\phi)\sqrt{1-v^2}\,.\end{equation}
Substituting the Lagrangian into the Euler--Lagrange equation and taking the non-relativistic limit gives
\begin{equation}\frac{d^2x^i}{dt^2}\approx-\kappa\alpha\left(\phi\right)\left(\partial_i\phi+\dot{\phi}\frac{dx^i}{dt}\right)\,.\end{equation}
Here, $\alpha = \frac{1}{\kappa^2 m(\phi)} \frac{\partial [\kappa m(\phi)]}{\partial \phi}$ characterizes the scalar coupling to the test mass~\cite{Damour2010,Will2014,Touboul2022}. For an initially stationary test mass, the second term is higher order in the coupling than the first and can be neglected at linear order. The scalar field then induces the acceleration
\begin{equation}
a^i\left(t,\vec{x}\right)=i\phi_{\vec{k}}\alpha\kappa k^ie^{im_\phi\left(t-v\hat{k}\cdot\vec{x}\right)}\,.
\end{equation}
Integrating the oscillatory acceleration twice gives the induced displacement,
\begin{equation}\delta x^i\left(t,\vec{x}\right)=-i\mathcal{M}_s{\hat{k}}^ie^{im_\phi\left(t-v\hat{k}\cdot\vec{x}\right)}\,,\end{equation}
where $\mathcal{M}_s=\alpha\kappa\phi_{\vec{k}}|\vec{k}|/m_\phi^2$ sets the displacement amplitude. The induced displacements of the sending and receiving test masses change the light-travel time and hence the measured laser frequency.

\subsection{Form of the response of the space-based detector to dark matter}

\begin{figure}[t]
    \centering
\begin{tikzpicture}[x=1cm,y=1cm,font=\fontsize{9}{11}\selectfont,
  >={Stealth[length=1.5mm]},line cap=round,line join=round]
  \coordinate (s1) at (0,0);
  \coordinate (s2) at (1.1,-2.8);
  \coordinate (s3) at (-2.8,-1.55);
  \draw[blue!65!cyan,line width=.7pt] (s1)--node[left=3pt] {$L_2$}(s3)
    --node[below=3pt] {$L_1$}(s2)--node[right=3pt] {$L_3$}(s1);
  \draw[->,line width=.5pt] (0,.2)--(0,1.55) node[above] {$z$};
  \draw[->,line width=.5pt] (.24,0)--(3.0,0) node[right] {$y$};
  \draw[->,line width=.5pt] (-.15,-.19)--(-2.2,-2.8) node[below left] {$x$};
  \draw ({.62*cos(232)},{.62*sin(232)}) arc[start angle=232,end angle=291.45,radius=.62];
  \node at ({.88*cos(262)},{.88*sin(262)}) {$\alpha$};
  \draw ({.91*cos(208.97)},{.91*sin(208.97)}) arc[start angle=208.97,end angle=232,radius=.91];
  \node at ({1.19*cos(220)},{1.19*sin(220)}) {$\beta$};
  \foreach \s in {s1,s2,s3}{
    \begin{scope}[shift={(\s)}]
      \filldraw[fill=gray!18,draw=black!75,line width=.4pt] (-.09,-.105) rectangle (.09,.105);
      \filldraw[fill=blue!20,draw=black!65,line width=.3pt] (-.28,-.085) rectangle (-.13,.085);
      \filldraw[fill=blue!20,draw=black!65,line width=.3pt] (.13,-.085) rectangle (.28,.085);
      \draw[black!65,line width=.3pt] (-.13,0)--(-.09,0) (.09,0)--(.13,0);
      \draw[blue!45!black,line width=.2pt] (-.205,-.085)--(-.205,.085) (.205,-.085)--(.205,.085);
    \end{scope}
  }
  \node[above right=2pt] at (s1) {SC1};
  \node[below right=2pt] at (s2) {SC2};
  \node[above left=2pt] at (s3) {SC3};
\end{tikzpicture}
    \caption{Schematic diagram of the geometry of the triangular space-based detector.}
    \label{fig:space-based detector}
\end{figure}
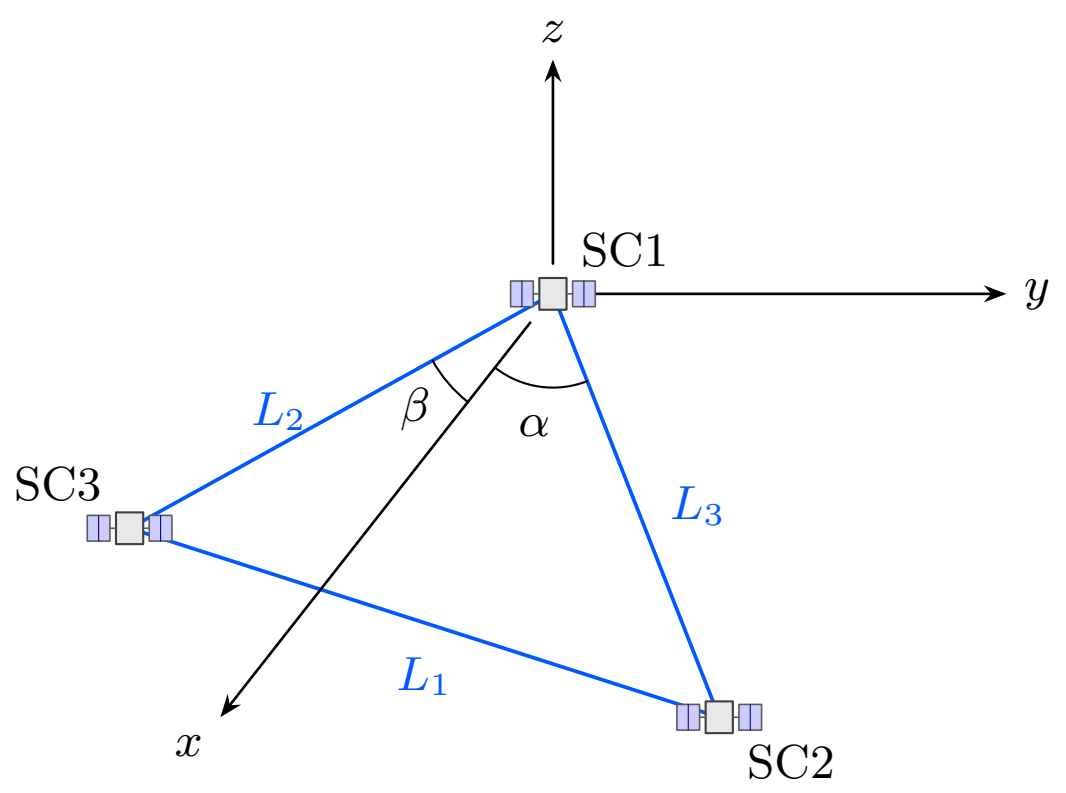

We now obtain the laser-frequency signal from the test-mass displacement, accounting for the light-travel time between spacecraft. We use the link geometry in figure~\ref{fig:space-based detector}. The spacecraft are labeled SC1, SC2, and SC3. Here $\alpha$ and $\beta$ are the geometric angles between the $x$ axis and the arms SC1--SC2 and SC1--SC3, respectively. The arm opposite SC$i$ has length $L_i$.
The position vectors of the three satellites are:
\begin{equation}
\begin{aligned}
\vec{x}_1 &= (0, 0, 0),\\
\vec{x}_2 &= L_3 (\cos(\alpha), \sin(\alpha), 0),\\
\vec{x}_3 &= L_2 (\cos(\beta), -\sin(\beta), 0)\,.
\end{aligned}
\end{equation}
Accordingly, the direction vectors along the arms are defined as:
\begin{equation}
\begin{aligned}
{\vec{n}}_{12}&=\frac{1}{L_3}\left({\vec{x}}_2-{\vec{x}}_1\right)=\left(\cos{\left(\alpha\right)},\sin{\left(\alpha\right)},0\right),\\
{\vec{n}}_{23}&=\frac{1}{L_1}\left({\vec{x}}_3-{\vec{x}}_2\right) \\
&\quad = \biggl(\frac{L_2\cos{\left(\beta\right)}-L_3\cos{\left(\alpha\right)}}{L_1}, \\
&\qquad \frac{-L_2\sin{\left(\beta\right)}-L_3\sin{\left(\alpha\right)}}{L_1},0\biggr),\\
{\vec{n}}_{31}&=\frac{1}{L_2}\left({\vec{x}}_1-{\vec{x}}_3\right)=\left(-\cos{\left(\beta\right)},\sin{\left(\beta\right)},0\right)\,.
\end{aligned}
\end{equation}
The angles $\theta$ and $\phi$ specify the field propagation direction in spherical coordinates. The corresponding unit vector $\hat{k}$ and wave vector $\vec{k}$ are
\begin{equation}\begin{aligned}
\hat{k}&=\left(\sin\theta\cos\phi,\sin\theta\sin\phi,\cos\theta\right),\\
\vec{k}&=\frac{2\pi f}{c}v\left(\sin\theta\cos\phi,\sin\theta\sin\phi,\cos\theta\right)\,.
\end{aligned}
\end{equation}
With the link directions specified, we substitute the induced displacements into the light-travel-time perturbation. For a link received at SC$r$ and emitted at SC$s$, the propagation time $T_{rs}$ is
\begin{equation}T_{rs}=L_{rs}+\delta t_{rs}\,,\end{equation}
where $L_{rs}$ is the unperturbed light-travel time in natural units and $\delta t_{rs}$ is its perturbation. The unit vector $\hat n_{rs}$ points from receiver to sender. In the equal-arm expressions below, the common delay is $L$, so the displacement perturbation is
\begin{equation}\delta t_{rs}=-{\hat{n}}_{rs}\cdot\left[\delta\vec{x}\left(t,{\vec{x}}_r\right)-\delta\vec{x}\left(t-L,{\vec{x}}_s\right)\right]\,.\end{equation}
This perturbation changes the received laser phase according to
\begin{equation}\phi_r\left(t\right)=\phi_s\left(t-T_{rs}\right)=2\pi\nu_0\left(t-L_{rs}-\delta t_{rs}\right)\,,\end{equation}
where $\nu_0$ is the reference laser frequency and $\phi_s$, $\phi_r$ denote the emitted and received laser phases. The received instantaneous frequency is
\begin{equation}\nu_{rs}\left(t\right)=\frac{1}{2\pi}\frac{d\phi_r}{dt}\,.\end{equation}
Writing $\delta\nu_{rs}=\nu_{rs}-\nu_0$, the phase derivative gives the relative frequency shift
\begin{equation}\frac{\delta\nu_{rs}}{\nu_0}=-\frac{d\delta t_{rs}}{dt}\,.\end{equation}
Substituting the test-mass displacement gives the single-link Doppler response in the unified form \cite{Estabrook1975}:
\begin{equation}
\label{eq:y}
y_{rs}\left(t\right)=\frac{\delta\nu_{rs}}{\nu_0}=\mu_{rs}\left[h\left(t,{\vec{x}}_r\right)-h\left(t-L,{\vec{x}}_s\right)\right]\,,
\end{equation}
where $h(t,\vec{x})=m_\phi\mathcal{M}_s e^{im_\phi(t-v\hat{k}\cdot\vec{x})}$ is the scalar-signal amplitude used to normalize the response and $\mu_{rs}=\hat k\cdot\hat n_{rs}$ is the projection along the link.
For the field at the origin, $h(t)\equiv h(t,\vec 0)$, we use the temporal Fourier convention
\begin{equation}h(t)=\frac{1}{2\pi}\int_{-\infty}^{\infty}\widetilde h(\omega)e^{i\omega t}\,d\omega.\end{equation}
A finite observation of duration $T$ is described by $\widetilde h_T(\omega)=\int_0^T h(t)e^{-i\omega t}\,dt$.
Substituting into Eq.~\eqref{eq:y} gives:
\begin{equation}
\begin{split}
y_{rs}\left(t\right)&=\mu_{rs}\frac{1}{2\pi}\int_{-\infty}^{\infty} d\omega\widetilde{h}\left(\omega\right)e^{i\omega t} \\
&\quad \times \left[e^{-i\vec{k}\cdot{\vec{x}}_r}-e^{-i\left(\tau+\vec{k}\cdot{\vec{x}}_s\right)}\right]\,,
\end{split}
\end{equation}
where $\tau=2\pi fL/c$. The response expression for a single link is then obtained as
\begin{equation}{\widetilde{y}}_{rs}\left(\omega\right)=\mu_{rs}\widetilde{h}\left(\omega\right)\left(e^{-i\left(\vec{k}\cdot{\vec{x}}_r\right)}-e^{-i\left(\tau+\vec{k}\cdot{\vec{x}}_s\right)}\right)\,,\end{equation}
where $\mu_{rs}=\hat{k}\cdot{\hat{n}}_{rs}$. This single-link frequency-domain response provides the input for constructing TDI observables from delayed one-way measurements.

\section{Response functions of dark matter under multiple types of TDI combinations}
\label{sec:3}

Starting from the single-link Doppler response derived above, we construct TDI observables by combining delayed measurements. In this section, we express the direction-averaged power response in terms of six delay coefficients and verify the analytic results for A, E, and T by numerical angular integration.

\subsection{Time-delay interferometry combinations}

\begin{figure*}[t]
\centering
\begin{tikzpicture}[x=1.25cm,y=1cm,font=\fontsize{9}{11}\selectfont,
  >={Stealth[length=1.4mm]},
  sc/.style={draw=black!65,fill=gray!7,rounded corners=1.2pt,inner sep=3pt}]
  \foreach \x/\sc [count=\i from 0] in {0/1,1.6/3,3.2/1,4.8/2,6.4/1}
    \node[sc] (a\i) at (\x,1.35) {SC\sc};
  \foreach \x/\sc [count=\i from 0] in {0/1,1.6/2,3.2/1,4.8/3,6.4/1}
    \node[sc] (b\i) at (\x,0) {SC\sc};
  \node[anchor=east] at (-.45,1.35) {path a};
  \node[anchor=east] at (-.45,0) {path b};
  \foreach \l/\r/\v in {0/1/{D^3y_{31}},1/2/{D^2y_{13}},2/3/{Dy_{21}},3/4/{y_{12}}}
    \draw[->,blue!65!black,line width=.65pt] (a\l)--node[above=3pt] {$\v$}(a\r);
  \foreach \l/\r/\v in {0/1/{D^3y_{21}},1/2/{D^2y_{12}},2/3/{Dy_{31}},3/4/{y_{13}}}
    \draw[->,orange!75!black,line width=.65pt] (b\l)--node[above=3pt] {$\v$}(b\r);
  \node[anchor=north] at (3.2,-.45) {$X=(\text{path b})-(\text{path a})$};
\end{tikzpicture}
\caption{The two synthesized light paths of the equal-arm Michelson $X$ combination. The labels show each one-way measurement and its delay relative to the final readout. SC1--SC3 denote the three spacecraft.}\label{fig:tdi_X_paths}
\end{figure*}
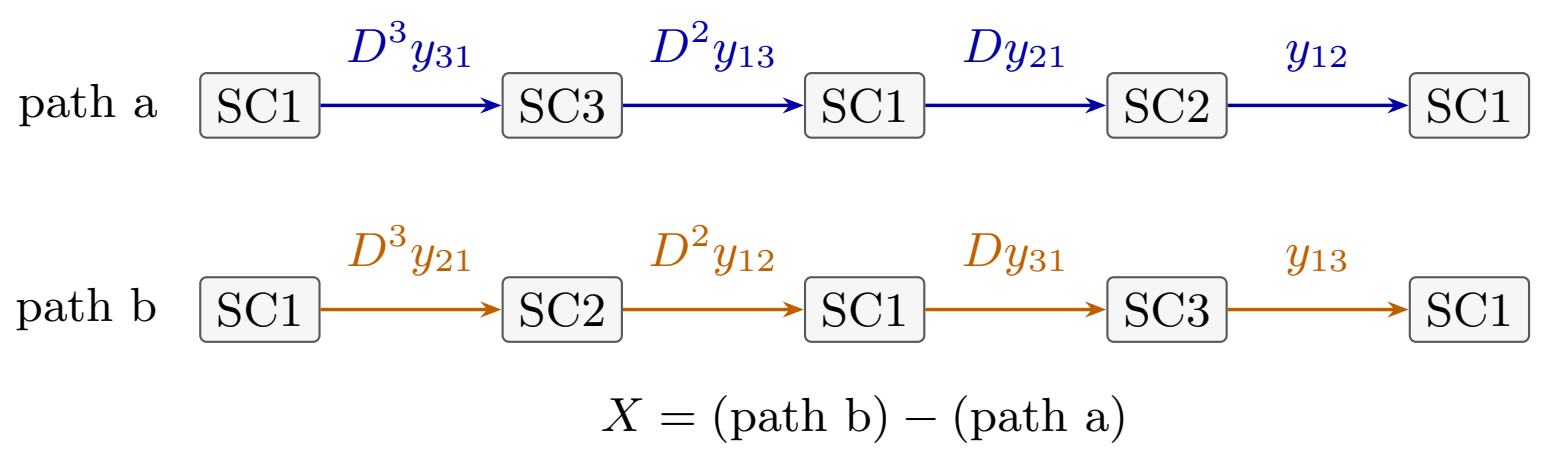

TDI observables suppress laser frequency noise~\cite{Tinto1999,Armstrong1999,Estabrook2000,Tinto2021}. The chosen delay model determines the order to which laser noise is cancelled. Any TDI combination is therefore completely characterized by a set of delay polynomials, namely the coefficients of the individual single-link responses and the corresponding time delays.
As an example, figure~\ref{fig:tdi_X_paths} shows the two synthesized light paths of the equal-arm Michelson $X$ combination. Each arrow contributes a one-way measurement. A one-arm delay is $D=e^{-iu}$, with $u=2\pi fL/c=\tau$. Each additional one-arm light-travel delay introduces another factor of $D$, so path b minus path a gives
\begin{equation}
X=(1-D^2)\left(y_{13}+Dy_{31}-y_{12}-Dy_{21}\right)\,.\label{eq:X_example}
\end{equation}

The commonly used elementary combinations belong to first-generation TDI. The Michelson combinations X, Y, and Z each take one spacecraft as the vertex and compare round trips along its two arms. The Sagnac combinations $\alpha$, $\beta$, and $\gamma$ correspond to closed loops around the three spacecraft. Other elementary combinations include the Beacon combinations P, Q, and R, the Monitor combinations E, F, and G, and the Relay combinations U, V, and W  \cite{Armstrong1999,Tinto2021}. Orthogonal linear combinations of X, Y, and Z give the combinations A, E, and T. Under symmetric noise assumptions these are noise-orthogonal; the T response is suppressed in the long-wavelength limit for gravitational waves, but it is not an exactly signal-free channel~\cite{Wang2023,Prince2002}. Under the equal-arm approximation, some differently named combinations have identical responses. In this paper we extend the elementary combinations above to geometric 12-link, 14-link, and 16-link TDI combinations \cite{Wang2023} and derive analytic expressions for their response functions. To treat these combinations in a common framework, we express the response in terms of their six delay-polynomial coefficients.

\subsection{Analytic calculation of the response function}
\label{sec:analytic_response}

In $y_{rs}$, $r,s\in\{1,2,3\}$ label the receiving and transmitting spacecraft, respectively, with $r\ne s$. We use the six-link order $(y_{12},y_{23},y_{31},y_{13},y_{21},y_{32})$ with coefficients $(P_1,P_2,P_3,P_{1\prime},P_{2\prime},P_{3\prime})$. We denote by $\widetilde{\mathrm{TDI}}(\omega)$ the TDI observable specified by these coefficients. Its direction-averaged power response is~\cite{Yu2023,Tinto2021,Wang2023}
\begin{equation}
R_{\mathrm{TDI}}(\omega) = \frac{1}{4\pi} \int_{-1}^{1} d\cos\theta \int_{0}^{2\pi} d\phi \left| \frac{\widetilde{\mathrm{TDI}}(\omega)}{\tilde{h}(\omega)} \right|^2\,.
\end{equation}
We call the corresponding quantity $\sqrt R$ the transfer function.
\begin{widetext}
Substituting the six single-link responses gives the squared modulus inside the integral as
\begin{equation}
\left|\begin{aligned}
&P_1\mu_{12}\left(e^{-i\vec k\cdot\vec x_1}-e^{-i(\tau+\vec k\cdot\vec x_2)}\right)
+P_2\mu_{23}\left(e^{-i\vec k\cdot\vec x_2}-e^{-i(\tau+\vec k\cdot\vec x_3)}\right)\\
&+P_3\mu_{31}\left(e^{-i\vec k\cdot\vec x_3}-e^{-i(\tau+\vec k\cdot\vec x_1)}\right)
+P_{1\prime}\mu_{13}\left(e^{-i\vec k\cdot\vec x_1}-e^{-i(\tau+\vec k\cdot\vec x_3)}\right)\\
&+P_{2\prime}\mu_{21}\left(e^{-i\vec k\cdot\vec x_2}-e^{-i(\tau+\vec k\cdot\vec x_1)}\right)
+P_{3\prime}\mu_{32}\left(e^{-i\vec k\cdot\vec x_3}-e^{-i(\tau+\vec k\cdot\vec x_2)}\right)
\end{aligned}\right|^2\,,
\end{equation}
and rearranged into the form
\begin{equation}R_{\mathrm{TDI}}(\omega)=\left\langle|A+B+C|^2\right\rangle_{\Omega},\qquad \langle\cdot\rangle_\Omega=\frac{1}{4\pi}\int d\Omega\,(\cdot).\end{equation}
Here $A$, $B$, and $C$ are given by
\begin{equation}
\begin{split}
A &= \mu_{12} \left( \left( P_1 + e^{-i\tau} P_{2^\prime} \right) e^{-i\vec{k}\cdot\vec{x}_1} - \left( P_1 e^{-i\tau} + P_{2^\prime} \right) e^{-i\vec{k}\cdot\vec{x}_2} \right), \\
B &= \mu_{23} \left( \left( P_2 + e^{-i\tau} P_{3^\prime} \right) e^{-i\vec{k}\cdot\vec{x}_2} - \left( P_2 e^{-i\tau} + P_{3^\prime} \right) e^{-i\vec{k}\cdot\vec{x}_3} \right), \\
C &= \mu_{31} \left( \left( P_3 + e^{-i\tau} P_{1^\prime} \right) e^{-i\vec{k}\cdot\vec{x}_3} - \left( P_3 e^{-i\tau} + P_{1^\prime} \right) e^{-i\vec{k}\cdot\vec{x}_1} \right)\,.
\end{split}
\end{equation}
For convenience we write out the expression for $\mu$
\begin{equation}
\begin{aligned}
\mu_{12} &= \hat{k} \cdot \hat{n}_{12} = \sin\theta \cos\phi \cos\alpha + \sin\theta \sin\phi \sin\alpha = \sin\theta \cos(\phi - \alpha),\\
\mu_{23} &= \hat{k} \cdot \hat{n}_{23} = \sin\theta \cos(\phi + \delta),\\
\mu_{31} &= \hat{k} \cdot \hat{n}_{31} = -\sin\theta \cos\phi \cos\beta + \sin\theta \sin\phi \sin\beta = -\sin\theta \cos(\phi + \beta)\,,
\end{aligned}
\end{equation}
Here $u_i\equiv2\pi fL_i/c$ is the optical propagation phase of arm $i$, with $u_i=u$ in the equal-arm limit. The angle $\delta$ satisfies $\tan\delta=(u_2\sin\beta+u_3\sin\alpha)/(u_2\cos\beta-u_3\cos\alpha)$.
The angular average can then be decomposed into six terms,
\begin{equation}R_{\mathrm{TDI}}(\omega)=\left\langle |A|^2+|B|^2+|C|^2+2\mathrm{Re}[AB^*]+2\mathrm{Re}[AC^*]+2\mathrm{Re}[BC^*]\right\rangle_\Omega\,.\label{eq:angular_response_terms}
\end{equation}
To express these angular integrals, we define the functions
\begin{equation}\begin{aligned}
f(u_i)&=\frac{2u_i\cos\left(u_i\right)+(-2+u_i^2)\sin(u_i)}{u_i^3},\\
g(u_i)&=\frac{u_i\cos\left(u_i\right)+(-1+u_i^2)\sin(u_i)}{u_i^3}\,.
\end{aligned}
\end{equation}
The spatial-phase arguments of these integrals are $vu_i$, while the optical delay phases are $u_i$. The functions have removable limits $f(0)=1/3$ and $g(0)=2/3$, and high precision is used near these limits. The symbol $\mathrm{Re}$ denotes the real part and the asterisk denotes complex conjugation.
The first three terms are the angular averages of the squared contributions from the individual arm pairs. Explicitly, we obtain:
\begin{equation}
\begin{split}
\langle|A|^2\rangle_\Omega &= \frac{1}{3} \left[ \left( P_1 + e^{-i\tau} P_{2^\prime} \right) \left( P_1^\ast + e^{i\tau} P_{2^\prime}^\ast \right) + \left( P_1 e^{-i\tau} + P_{2^\prime} \right) \left( P_1^\ast e^{i\tau} + P_{2^\prime}^\ast \right) \right] \\
&\quad - 2 f(vu_3) \mathrm{Re} \left[ \left( P_1 + e^{-i\tau} P_{2^\prime} \right) \left( P_1^\ast e^{i\tau} + P_{2^\prime}^\ast \right) \right]\,.
\end{split}
\end{equation}
\begin{equation}
\begin{split}
\langle|B|^2\rangle_\Omega &= \frac{1}{3} \left[ (P_2 + e^{-i\tau} P_{3^\prime})(P_2^\ast + e^{i\tau} P_{3^\prime}^\ast) + (P_2 e^{-i\tau} + P_{3^\prime})(P_2^\ast e^{i\tau} + P_{3^\prime}^\ast) \right] \\
&\quad - 2 f(vu_1) \mathrm{Re} \left[ (P_2 + e^{-i\tau} P_{3^\prime})(P_2^\ast e^{i\tau} + P_{3^\prime}^\ast) \right]\,.
\end{split}
\end{equation}
\begin{equation}
\begin{split}
\langle|C|^2\rangle_\Omega &= \frac{1}{3} \left[ (P_3 + e^{-i\tau} P_{1^\prime})(P_3^\ast + e^{i\tau} P_{1^\prime}^\ast) + (P_3 e^{-i\tau} + P_{1^\prime})(P_3^\ast e^{i\tau} + P_{1^\prime}^\ast) \right] \\
&\quad - 2 f(vu_2) \mathrm{Re} \left[ (P_3 + e^{-i\tau} P_{1^\prime})(P_3^\ast e^{i\tau} + P_{1^\prime}^\ast) \right]\,.
\end{split}
\end{equation}
The three squared terms have the same form under cyclic relabeling of the arms and their link coefficients. Each contains the powers of the two endpoint contributions and an interference term weighted by $f(vu_i)$ for arm $i$. The remaining three terms describe interference between different arm pairs that survives direction averaging. These cross terms are given by
\begin{equation}
\begin{split}
2\mathrm{Re}\langle AB^*\rangle_\Omega = & -\frac{1}{3}\cos(\alpha + \delta) \Big[ (P_1 e^{-i\tau} + P_{2'})(P_2^* + e^{i\tau}P_{3'}^*) \\
& + (P_1^* e^{i\tau} + P_{2'}^*)(P_2 + e^{-i\tau}P_{3'}) \Big] \\
& + 2\cos(\alpha + \delta)f(vu_3)\mathrm{Re} \Big[ (P_1 + e^{-i\tau}P_{2'})(P_2^* + e^{i\tau}P_{3'}^*) \Big] \\
& + \Big[ -2\cos(\alpha + 2\beta - \delta)f(vu_2) \\
& + 2\sin(\alpha + \beta)\sin(\delta - \beta)g(vu_2) \Big] \mathrm{Re} \Big[ (P_1 + e^{-i\tau}P_{2'})(P_2^* e^{i\tau} + P_{3'}^*) \Big] \\
& + 2\cos(\alpha + \delta)f(vu_1) \mathrm{Re} \Big[ (P_1 e^{-i\tau} + P_{2'})(P_2^* e^{i\tau} + P_{3'}^*) \Big]\,.
\end{split}
\end{equation}
\begin{equation}
\begin{split}
2\mathrm{Re}\langle BC^*\rangle_\Omega = & \frac{1}{3}\cos(\delta - \beta) \Big[ (P_2 e^{-i\tau} + P_{3'})(P_3^* + e^{i\tau}P_{1'}^*) \\
& + (P_2^* e^{i\tau} + P_{3'}^*)(P_3 + e^{-i\tau}P_{1'}) \Big] \\
& - 2\cos(\delta - \beta)f(vu_1)\mathrm{Re} \Big[ (P_2 + e^{-i\tau}P_{3'})(P_3^* + e^{i\tau}P_{1'}^*) \Big] \\
& + \Big[ 2\cos(2\alpha + \beta + \delta)f(vu_3) \\
& + 2\sin(\alpha + \delta)\sin(\alpha + \beta)g(vu_3) \Big] \mathrm{Re} \Big[ (P_2 + e^{-i\tau}P_{3'})(P_3^* e^{i\tau} + P_{1'}^*) \Big] \\
& - 2\cos(\delta - \beta)f(vu_2)\mathrm{Re} \Big[ (P_2 e^{-i\tau} + P_{3'})(P_3^* e^{i\tau} + P_{1'}^*) \Big]\,. 
\end{split}
\end{equation}
\begin{equation}
\begin{split}
2\mathrm{Re}\langle AC^*\rangle_\Omega = & \frac{1}{3}\cos(\alpha + \beta) \Big[ (P_1 + e^{-i\tau}P_{2'})(P_3^* e^{i\tau} + P_{1'}^*) \\
& + (P_1^* + e^{i\tau}P_{2'}^*)(P_3 e^{-i\tau} + P_{1'}) \Big] \\
& - 2\cos(\alpha + \beta)f(vu_2)\mathrm{Re} \Big[ (P_1 + e^{-i\tau}P_{2'})(P_3^* + e^{i\tau}P_{1'}^*) \Big] \\
& + \Big[ 2\cos(\alpha - \beta + 2\delta)f(vu_1) \\
& - 2\sin(\alpha + \delta)\sin(\beta - \delta)g(vu_1) \Big] \mathrm{Re} \Big[ (P_1 e^{-i\tau} + P_{2'})(P_3^* + e^{i\tau}P_{1'}^*) \Big] \\
& - 2\cos(\alpha + \beta)f(vu_3)\mathrm{Re} \Big[ (P_1 e^{-i\tau} + P_{2'})(P_3^* e^{i\tau} + P_{1'}^*) \Big]\,.
\end{split}
\end{equation}
\end{widetext}

The interference terms between different arm pairs can be positive or negative, depending on the geometry and the relative phases. Similar squared terms can therefore give different total responses.

For fixed detector geometry and field parameters, the same angular kernels apply to all the combinations considered here. Each combination supplies its own six delay coefficients, and their contributions sum to the direction-averaged power response in Eq.~\eqref{eq:angular_response_terms}. Since all six terms are quadratic in these coefficients, a common factor $q(D)$ scales the power response by $|q(D)|^2$. Section~\ref{sec:4} uses this property to compare the sensitivities of different combinations.

\subsection{Numerical verification}

To verify the analytic results derived above, we compare them with numerical angular integrals of the corresponding responses. As concrete examples, we use the optimal combinations A, E, and T formed from the Michelson combinations X, Y, and Z in an equal-arm, equilateral-triangle detector \cite{Yu2023,Tinto2021,Wang2023,Prince2002}. In the time domain, these combinations are defined as follows:
\begin{equation}A\left(t\right)=\frac{1}{\sqrt2}\left[Z\left(t\right)-X\left(t\right)\right]\,.\end{equation}
\begin{equation}E\left(t\right)=\frac{1}{\sqrt6}\left[X\left(t\right)-2Y\left(t\right)+Z\left(t\right)\right]\,.\end{equation}
\begin{equation}T\left(t\right)=\frac{1}{\sqrt3}\left[X\left(t\right)+Y\left(t\right)+Z\left(t\right)\right]\,.\end{equation}
For A, E, and T, the six-link coefficients follow from the delay polynomials of X, Y, and Z. Their response functions are given below:
\begin{equation}R_A(\omega)=\left(1-\cos{2\tau}\right)\left[Q_1^2+Q_2^2\right]\,.\end{equation}
\begin{equation}R_E(\omega)=\frac{1}{3}\left(1-\cos{2\tau}\right)\left[Q_3^2+Q_4^2\right]\,.\end{equation}
\begin{equation}R_T(\omega)=\frac{64}{3}\sin^4({\frac{\tau}{2}})\sin^2{\tau}\left[Q_5^2+Q_6^2\right]\,.\end{equation}
The functions $Q_{1}$ and $Q_{2}$ correspond to the combination A:
\begin{equation}
\begin{split}
Q_1&=- (\mu_{32} + \mu_{12}) [1 + \cos(2\tau)] \\
&\quad + 2\mu_{32} \cos(\tau + \vec{k} \cdot \vec{x}_2 + \vec{k} \cdot \vec{x}_3) \\
&\quad + 2\mu_{12} \cos(\tau + \vec{k} \cdot \vec{x}_1 + \vec{k} \cdot \vec{x}_2)\,.
\end{split}
\end{equation}
\begin{equation}
\begin{split}
Q_2&=(\mu_{32} + \mu_{12}) \sin(2\tau) \\
&\quad - 2\mu_{32} \sin(\tau + \vec{k} \cdot \vec{x}_2 + \vec{k} \cdot \vec{x}_3) \\
&\quad - 2\mu_{12} \sin(\tau + \vec{k} \cdot \vec{x}_1 + \vec{k} \cdot \vec{x}_2)\,.
\end{split}
\end{equation}
The functions $Q_{3}$ and $Q_{4}$ correspond to the combination E:
\begin{equation}
\begin{split}
Q_3 & = -(2\mu_{13} + \mu_{23} - \mu_{21})[1 + \cos(2\tau)] \\
&\quad + 4\mu_{13} \cos(\tau + \vec{k} \cdot \vec{x}_1 + \vec{k} \cdot \vec{x}_3) \\
&\quad + 2\mu_{23} \cos(\tau + \vec{k} \cdot \vec{x}_2 + \vec{k} \cdot \vec{x}_3) \\
&\quad - 2\mu_{21} \cos(\tau + \vec{k} \cdot \vec{x}_1 + \vec{k} \cdot \vec{x}_2)\,. 
\end{split}
\end{equation}
\begin{equation}
\begin{split}
Q_4 & = (2\mu_{13} + \mu_{23} - \mu_{21}) \sin(2\tau) \\
&\quad - 4\mu_{13} \sin(\tau + \vec{k} \cdot \vec{x}_1 + \vec{k} \cdot \vec{x}_3) \\
&\quad - 2\mu_{23} \sin(\tau + \vec{k} \cdot \vec{x}_2 + \vec{k} \cdot \vec{x}_3) \\
&\quad + 2\mu_{21} \sin(\tau + \vec{k} \cdot \vec{x}_1 + \vec{k} \cdot \vec{x}_2)\,.
\end{split}
\end{equation}
The functions $Q_{5}$ and $Q_{6}$ correspond to the combination T:
\begin{equation}
\begin{split}
Q_5&=\mu_{13}[1 + \cos(\vec{k} \cdot \vec{x}_3)] \\
&\quad + \mu_{21}[\cos(\vec{k} \cdot \vec{x}_2) + 1] \\
&\quad - \mu_{23}[\cos(\vec{k} \cdot \vec{x}_2) + \cos(\vec{k} \cdot \vec{x}_3)]\,.
\end{split}
\end{equation}
\begin{equation}
\begin{split}
Q_6&=\mu_{13} \sin(\vec{k} \cdot \vec{x}_3) \\
&\quad + \mu_{21} \sin(\vec{k} \cdot \vec{x}_2) \\
&\quad - \mu_{23} [\sin(\vec{k} \cdot \vec{x}_2) + \sin(\vec{k} \cdot \vec{x}_3)]\,.
\end{split}
\end{equation}

Under the same equal-arm assumptions, we integrate the directional power responses given by $Q_1$--$Q_6$ numerically using $10^5$ isotropically sampled directions at frequencies from $10^{-5}$ to $1\,\mathrm{Hz}$. We compare these integrals with the analytic results derived in section~\ref{sec:analytic_response}.
To quantify this comparison, we report the relative amplitude deviation $\epsilon_{\sqrt R}=\sqrt{R_{\rm MC}/R_{\rm an}}-1$ and the normalized power deviation $z=(R_{\rm MC}-R_{\rm an})/\sigma_{\rm MC}$. Here $p_j$ is the unaveraged power response constructed from the auxiliary functions at the $j$th independent isotropic sample. The Monte Carlo estimate and its standard error are
\begin{equation}
\begin{gathered}
R_{\rm MC}=\frac{1}{M}\sum_{j=1}^{M}p_j, \\
\sigma_{\rm MC}=\left[\frac{1}{M(M-1)}\sum_{j=1}^{M}(p_j-R_{\rm MC})^2\right]^{1/2},
\end{gathered}
\label{eq:MC_standard_error}
\end{equation}
where $M=10^5$. Thus $\sigma_{\rm MC}$ estimates the standard error of the Monte Carlo mean power response. In the error panels, $\sigma$ instead denotes the standard error of $100\epsilon_{\sqrt R}$ under first-order propagation evaluated at $R_{\rm an}$:
\begin{equation}
\sigma=\frac{100\,\sigma_{\rm MC}}{2R_{\rm an}}.
\label{eq:amplitude_standard_error}
\end{equation}
Here $\sigma$ and $3\sigma$ are expressed in percentage points and are distinct from $\sigma_{\rm MC}$. Exact response zeros are excluded from relative-error statistics.

\begin{table}[t]\centering
\begin{tabular}{ccc}\hline
Combination & $\max|\epsilon_{\sqrt R}|$ & $\max|z|$\\\hline
A & $0.2257\%$ & 1.4796\\
E & $0.0564\%$ & 0.3993\\
T & $0.2621\%$ & 1.3779\\\hline
\end{tabular}
\caption{Comparison of analytic responses with Monte Carlo angular integrals of the expressions in terms of the auxiliary functions.}\label{tab:validation}
\end{table}

In the examined band and sampling configuration, the analytic responses agree well with the Monte Carlo angular integrals for A, E, and T. Their relative amplitude deviations are below approximately $0.3\%$, and every valid frequency satisfies $\lvert z\rvert\le 3$. This agreement confirms the correctness of the analytic angular averages for A, E, and T within the precision of the numerical comparison.

Figures~\ref{fig:A}--\ref{fig:T} compare the numerical and analytic results. Table~\ref{tab:validation} summarizes this comparison, and the right panels show the corresponding relative amplitude deviations in percent with the pointwise $\pm3\sigma$ ranges defined in Eq.~\eqref{eq:amplitude_standard_error}.
\begin{figure*}[p]\centering
\begin{minipage}[c]{0.55\linewidth}\centering
\input{figures/A_response_panel.tex}
\end{minipage}\hspace{0.02\linewidth}
\begin{minipage}[c]{0.34\linewidth}\centering
\input{figures/A_error_panel.tex}
\end{minipage}
\caption{Transfer function $\sqrt R$ of the A combination (left) and relative amplitude deviation $\epsilon_{\sqrt R}$ between numerical and analytic results (right).}\label{fig:A}
\par\smallskip
\begin{minipage}[c]{0.55\linewidth}\centering
\input{figures/E_response_panel.tex}
\end{minipage}\hspace{0.02\linewidth}
\begin{minipage}[c]{0.34\linewidth}\centering
\input{figures/E_error_panel.tex}
\end{minipage}
\caption{Transfer function $\sqrt R$ of the E combination (left) and relative amplitude deviation $\epsilon_{\sqrt R}$ between numerical and analytic results (right).}\label{fig:E}
\par\smallskip
\begin{minipage}[c]{0.55\linewidth}\centering
\input{figures/T_response_panel.tex}
\end{minipage}\hspace{0.02\linewidth}
\begin{minipage}[c]{0.34\linewidth}\centering
\input{figures/T_error_panel.tex}
\end{minipage}
\caption{Transfer function $\sqrt R$ of the T combination (left) and relative amplitude deviation $\epsilon_{\sqrt R}$ between numerical and analytic results (right).}\label{fig:T}
\end{figure*}

\section{Discussion}
\label{sec:4}

Following the numerical verification for A, E, and T, we compare the responses and sensitivities of the 12-link, 14-link, and 16-link TDI combinations under the same static equal-arm model~\cite{Wang2023}. We use their low-frequency behavior to classify the responses and explain how the sensitivities enter estimates of the scalar coupling.

For this comparison, let \(N_O(f)\) be the noise power spectral density (PSD) of a combination $O$. Its sensitivity curve is defined as
\begin{equation}S(f)=\frac{N_O(f)}{R_O(f)}\,,\end{equation}
where \(R_O(f)\) is the power response given in this paper. Thus $S$ is an equivalent input PSD, and $\sqrt S$ is the amplitude sensitivity in $\mathrm{Hz}^{-1/2}$. Once the laser noise is suppressed by TDI, two secondary noises limit detector performance: the optical metrology system noise, which dominates at high frequencies, and the test-mass acceleration noise, which dominates at low frequencies. 
We use LISA-like secondary-noise spectra~\cite{AmaroSeoane2017,Yu2023}. The optical metrology system (OMS) noise is given by:
\begin{equation}S_{\mathrm{oms}}\left(f\right)=\left(s_{\mathrm{oms}}\frac{2\pi f}{c}\right)^2\left[1+\left(\frac{2\times10^{-3}\mathrm{Hz} }{f}\right)^4\right]\frac{1}{\mathrm{Hz}}\,.\end{equation}
The test-mass acceleration (ACC) noise is given by:
\begin{equation}
\begin{split}
S_{\mathrm{acc}}\left(f\right)&=\left(\frac{s_{\mathrm{acc}}}{2\pi fc}\right)^2\left[1+\left(\frac{0.4\times10^{-3}\mathrm{Hz} }{f}\right)^2\right] \\
&\quad \times \left[1+\left(\frac{f}{8\times10^{-3}\mathrm{Hz} }\right)^4\right]\frac{1}{\mathrm{Hz}}\,.
\end{split}
\end{equation}
Assuming independent OMS and test-mass noise sources with identical spectra for all links and test masses, respectively, the six-link coefficients give the total noise PSD as~\cite{Wang2023}
\begin{equation}\label{eq:tdi_noise}
\begin{split}
N_O(f)={}&\left[S_{\mathrm{oms}}(f)+2S_{\mathrm{acc}}(f)\right]
\sum_{i=1}^{3}\left(|P_i|^2+|P_{i\prime}|^2\right)\\
&+4S_{\mathrm{acc}}(f)\cos u\,\mathrm{Re}\!\left[
P_1P_{2\prime}^{\ast}+P_2P_{3\prime}^{\ast}+P_3P_{1\prime}^{\ast}\right].
\end{split}
\end{equation}
Here the coefficients $P_i$ and $P_{i\prime}$ use the six-link order defined in section~\ref{sec:analytic_response}. The coefficients $s_{\mathrm{oms}}$ and $s_{\mathrm{acc}}$ set the optical displacement and test-mass acceleration noise amplitudes, respectively. The benchmark uses $L=3\times10^9\,\mathrm m$, $c=3\times10^8\,\mathrm{m\,s^{-1}}$, $v=10^{-3}$, $s_{\mathrm{oms}}=15\times10^{-12}\,\mathrm m$, and $s_{\mathrm{acc}}=3\times10^{-15}\,\mathrm{m\,s^{-2}}$.
\begin{figure*}[t]
    \centering
    \includegraphics[width=0.48\linewidth]{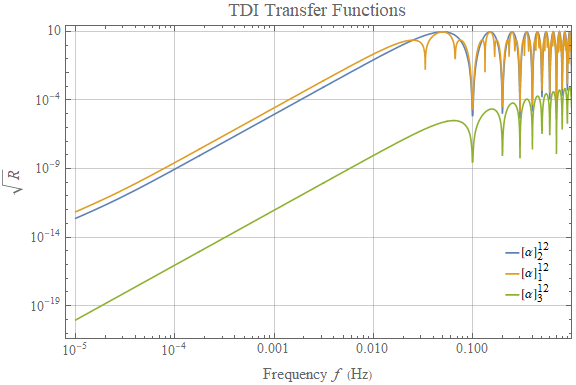} \includegraphics[width=0.48\linewidth]{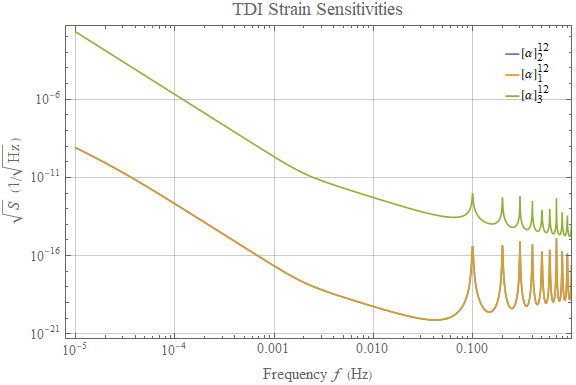}
    \caption{Transfer functions and sensitivity curves for the 12-link combinations.}
    \label{fig:12link}
\par\medskip
    \includegraphics[width=0.48\linewidth]{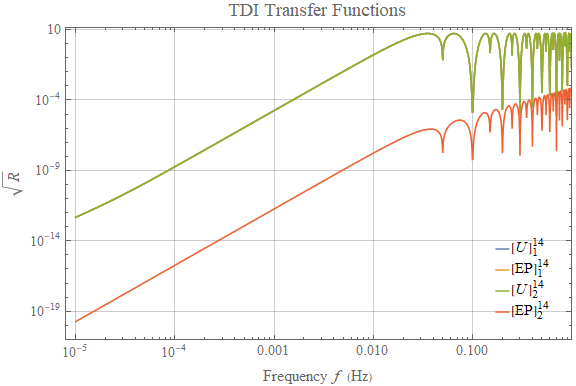} \includegraphics[width=0.48\linewidth]{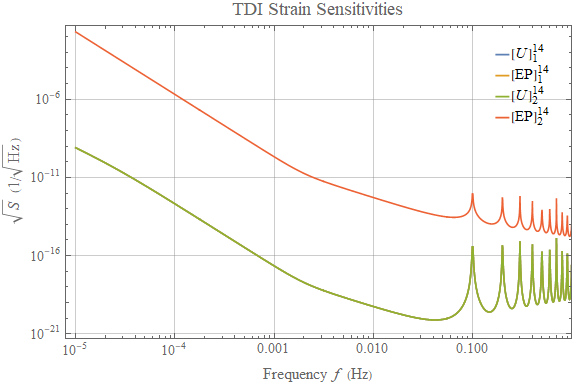}
    \caption{Transfer functions and sensitivity curves for the 14-link combinations.}
    \label{fig:14link}
\par\medskip
    \includegraphics[width=0.48\linewidth]{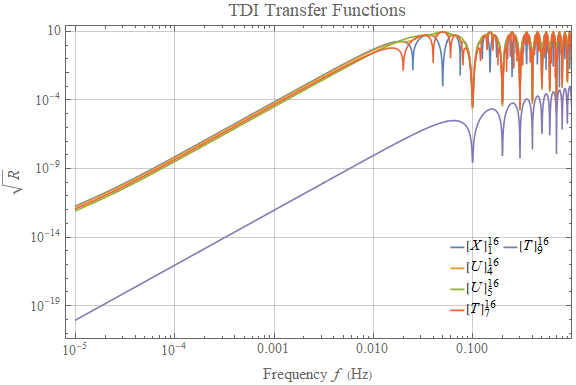} \includegraphics[width=0.48\linewidth]{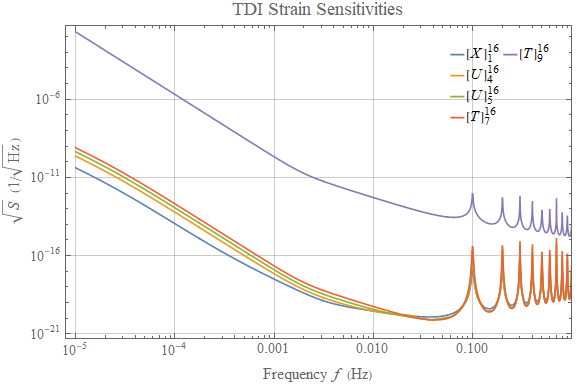}
    \caption{Transfer functions and sensitivity curves for the 16-link combinations.}
    \label{fig:16link}
\end{figure*}

The sets of 12-link, 14-link, and 16-link combinations contain 3, 4, and 38 combinations, respectively. The 12-link set contains two ordinary combinations, $[\alpha]_1^{12}$ and $[\alpha]_2^{12}$, and one $\zeta$ combination, $[\alpha]_3^{12}$. The 14-link ordinary combinations are $[U]_1^{14}$, $[U]_2^{14}$, and $[EP]_1^{14}$; their responses and noise spectra coincide in the equal-arm limit. The remaining combination, $[EP]_2^{14}$, is of $\zeta$ type. Among the 38 16-link combinations, six are of $\zeta$ type: $[T]_9^{16}$, $[T]_{12}^{16}$, $[T]_{14}^{16}$, $[T]_{15}^{16}$, $[T]_{18}^{16}$, and $[T]_{19}^{16}$. The remaining 32 are ordinary combinations.

To interpret the low-frequency differences between these combinations, we expand their responses for $u\ll1$ and $v\ll1$. The leading terms are
\begin{equation}
\begin{aligned}
R_n&=A_n\left(\frac4{15}v^2u^6+\frac19u^8\right)+O(v^2u^8)+O(u^{10}),\\
R_\zeta&=\frac{k}{280}v^4u^8+O(u^{10})\,.
\end{aligned}\label{eq:lowf}
\end{equation}
Here $A_n$ and $k$ are combination-dependent low-frequency coefficients. Ordinary combinations therefore have $R\propto v^2f^6$ for $u\ll v$, and $R\propto f^8$ for $v\ll u\ll1$. Their transfer functions $\sqrt R$ scale as $f^3$ and $f^4$, respectively. The $\zeta$ response lacks both leading ordinary terms and is suppressed by $v^4$ in the displayed low-frequency expansion. It does not exhibit the same two-term crossover. For example, $A_n=3$ for $[\alpha]_2^{12}$ and $A_n=9$ for $[U]_3^{16}$, while $k=1$ for $[\alpha]_3^{12}$ and $k=4$ for $[EP]_2^{14}$.

These expansions describe the low-frequency response, while the full response and noise spectra determine sensitivity across the band. Equal leading coefficients do not imply equal full response curves. Across all 45 combinations, the current equal-arm response and noise model gives 21 distinct power-response curves and 11 distinct single-channel amplitude-sensitivity curves. The latter are not 11 statistically independent channels. If two coefficient vectors satisfy $\boldsymbol P_b=q(D)\boldsymbol P_a$, then $R_b=|q|^2R_a$ and $N_b=|q|^2N_a$, so $S_b=S_a$ wherever the ratio is defined. A common filter zero must be treated by a limiting expression, rather than by interpreting $0/0$ as a sensitivity improvement or loss. For the 12-link ordinary pair, $q=1+D+D^2$; the larger low-frequency response of $[\alpha]_1^{12}$ is accompanied by proportionally larger noise.

To illustrate these comparisons, we select five representative 16-link combinations, shown in figure~\ref{fig:16link} and listed in table~\ref{tab:representatives}. Four ordinary combinations illustrate the frequency-dependent balance between signal and noise, and one represents the suppressed $\zeta$ response. On the verified finite grid from $10^{-4}\,\mathrm{Hz}$ to $0.1\,\mathrm{Hz}$, the amplitude-sensitivity envelope of the four ordinary representatives exceeds the full ordinary-channel optimum by at most approximately $3.09\%$.
\begin{table}[t]\centering
\begin{tabular}{ccc}\hline
Program index & Combination & Coefficient\\\hline
6 & $[X]_1^{16}$ & $A_n=192$\\
2 & $[U]_4^{16}$ & $A_n=147$\\
4 & $[U]_5^{16}$ & $A_n=39$\\
20 & $[T]_7^{16}$ & $A_n=75$\\
31 & $[T]_9^{16}$ & $k=1$\\\hline
\end{tabular}
\caption{Five displayed 16-link representatives. Columns give the program index, the combination, and the low-frequency coefficient in Eq.~\eqref{eq:lowf}. The program index refers to the enumeration used in the calculation files.}\label{tab:representatives}
\end{table}

The two leading terms in the ordinary response in Eq.~\eqref{eq:lowf} also locate the transition between the low-frequency regimes. Equating them gives the critical frequency
\begin{equation}f_c\simeq\sqrt{\frac{12}{5}}\frac{v_c}{2\pi L}.\end{equation}
Here $v_c\sim10^{-3}c\simeq3\times10^5\,\mathrm{m\,s^{-1}}$ is the field velocity. For the benchmark, the critical frequency is $2.47\times10^{-5}\,\mathrm{Hz}$, within the plotting range. The lowest plotted frequency is not deep in the $u\ll v$ regime. These power laws depend on the observable and its delay factors; they are not, by themselves, unique discriminators of scalar DM against gravitational-wave signals. Channel selection requires comparing $N/R$: a larger response or a larger link count does not alone imply better sensitivity. A suppressed scalar response may support a consistency check, but it does not guarantee a pure-noise channel or smaller noise.

The equivalent input sensitivity can be converted into a scalar-coupling sensitivity for space-based detectors through the effective scalar coupling~\cite{Yu2023},
\begin{equation}\alpha(\phi)\simeq d_g^\ast\simeq d_g+0.093(d_{\hat m}-d_g).\end{equation}
We assume that the field constitutes the local DM density $\rho_{\rm DM}$ and that the observation time $T$ is shorter than the field coherence time. With $\phi_{\vec k}=\sqrt{2\rho_{\rm DM}}/m_\phi$, the illustrative detectable-amplitude condition is
\begin{equation}
|\widetilde h(m_\phi)|=m_\phi\mathcal M_s\sqrt T
=d_g^\ast\kappa\phi_{\vec k}\frac{|\vec k|}{m_\phi}\sqrt T
=\sqrt{S_O},\label{eq.42}
\end{equation}
and hence
\begin{equation}d_g^\ast=\frac{m_\phi}{\kappa v_{\rm DM}}\sqrt{\frac{S_O}{2\rho_{\rm DM}T}}.\end{equation}
Here $S_O=N_O/R_O$. In Eq.~\eqref{eq.42}, $\widetilde h$ denotes the finite-time spectral amplitude $\widetilde h_T/\sqrt T$, using the observation window defined in section~\ref{sec:2}. The magnitude of the field velocity is $v_{\rm DM}\simeq|\vec k|/m_\phi\simeq v$. If $d_{\hat m}=0$, the quoted approximation gives $d_g^\ast\simeq0.907d_g$. This relation illustrates how the analytic responses obtained here enter projected coupling estimates. Such an estimate requires a specified local-field model, noise spectrum, integration prescription, and detection threshold. We do not derive a new confidence-level exclusion curve in this paper. The range $10^{-5}\,\mathrm{Hz}$--$1\,\mathrm{Hz}$ corresponds to rest energies $m_\phi c^2\simeq4.14\times10^{-20}\,\mathrm{eV}$--$4.14\times10^{-15}\,\mathrm{eV}$. Comparisons with fifth-force, equivalence-principle, or interferometric constraints require matching these assumptions~\cite{Yu2023,Touboul2022,Fukusumi2023,Miller2023}.

These response and sensitivity comparisons assume a static detector, a monochromatic plane wave, and a fixed field velocity $v$, and do not take spacecraft motion or orbital modulation into account. The equal-arm calculation is a baseline approximation. Although actual arm-length variations may be relatively small, their effect depends on frequency, required accuracy, and the observable. Near response zeros and in suppressed channels, relative changes can be large even for small arm variations. Time-dependent unequal arms also require a suitable TDI delay construction to maintain laser-noise cancellation; this requirement is distinct from the accuracy of the equal arm scalar-response approximation. The impact of these approximations on the sensitivity estimate can be quantified term by term in future work using the specific mission orbits.

\section{Conclusions}
\label{sec:5}

For ultralight scalar-field dark matter we have derived analytic expressions for the response functions of space-based gravitational-wave detectors to different TDI combinations in a static model, with the explicit comparisons performed in the equal-arm limit. We established a derivation chain from the scalar-field coupling action, through the test-mass displacement and the single-link Doppler response, to the response of a TDI combination. We give the general form of the response function for an arbitrary combination. For A, E, and T, numerical integration of the directional responses confirms the analytic angular averages within the precision of the comparison. These analytic expressions can be used to compare the response characteristics of different combinations and compute their sensitivity curves. The 45 geometric combinations are separated into ordinary and $\zeta$ types. The ordinary responses exhibit two low-frequency regimes, and the adopted noise model gives 11 distinct single-channel amplitude-sensitivity curves. Common signal and noise filtering explains why different responses can give the same $S=N/R$. Similar response functions have also been used to constrain scalar couplings with space-based gravitational-wave detectors \cite{Yu2023}. Confidence-level limits require separate statistical analysis. The above conclusions rest on the equal-arm, static-detector, and single-plane-wave assumptions. For more general orbital configurations and second-generation TDI, the response functions must be extended accordingly.

\begin{acknowledgments}
\label{sec:6}
This work was supported by the National Key R\&D
Program of China under Grant No. 2024YFC2207500
and the National Natural Science Foundation of China
under Grant No. 12405060.

\end{acknowledgments}

\section*{Data availability}

No new observational or experimental data were generated in this study.


\end{document}